\documentclass[useAMS,usenatbib,usegraphicx]{mn2e}

\newcommand{\etal}{et al.}
\newcommand{\mcg}{MCG--6-30-15}
\newcommand{\h}{1H\,0707--495}

\title[X-ray reflection in the NLS1 1H\,0707-495]
  {X-ray reflection in the NLS1 1H\,0707-495}
\author[A.\ C.\ Fabian \etal]
  {A.~C.~Fabian$^1$\thanks{acf@ast.cam.ac.uk}, G.~Miniutti$^1$,
  L.~Gallo$^2$, Th.~Boller$^2$, Y.~Tanaka$^2$, S.~Vaughan$^3$ and  
  R.R. Ross$^4$\\
  $^1$Institute of Astronomy, Madingley Road, Cambridge CB3 0HA \\
  $^2$Max-Planck-Institut f\"{u}r extraterrestrische Physik, Postfach
  1603, 85748 Garching, Germany\\
  $^3$Department of Physics and Astronomy, University of Leicester,
  University Road, Leicester LE1 7RH\\
  $^4$Physics Department, College of the Holy Cross, Worcester, MA
  01610, USA}

\pagerange{\pageref{firstpage}--\pageref{lastpage}}
\pubyear{2001}

\usepackage{times}

\begin{document}

\label{firstpage}

 \maketitle

\begin{abstract}
  We apply a reflection--dominated model to the second XMM-Newton
  observation of the Narrow Line Seyfert 1 galaxy \h. As in the first
  XMM-Newton observation a sharp spectral drop is detected with energy
  that has shifted from 7~keV to 7.5~keV in two years. The drop is
  interpreted in terms of relativistically blurred ionised reflection
  from the accretion disc, while the energy shift can be accounted for
  by changes in the ionisation state and, more importantly, emissivity
  profile on the disc. A flatter emissivity profile during the second
  higher flux observation reduces gravitational redshift effects,
  therefore shifting the edge to higher energy. Remarkably, ionised
  disc reflection and the associated power law continuum provide a
  reasonable description of the broadband spectrum, including the soft
  excess. Considering both observations, the spectral variability in
  \h\ appears to be due to the interplay between these two spectral
  components. The reflection component in the second observation is
  significantly less variable than the power law. Changes of the
  emissivity profile, spectral shape and variability properties (such
  as the rms spectrum) within the two observations are all consistent
  with a recently proposed model in which relativistic effects in the
  very inner regions of the nucleus play a major role.
\end{abstract}


\section{Introduction}
\label{sect:intro}

The first (hereafter GT for Guaranteed Time) XMM-Newton X-ray spectrum
of the Narrow Line Seyfert 1 galaxy, 1H\,0707-495 ($z=0.0411$), showed
a dramatic spectral drop at 7~keV (Boller et al 2002). This was
interpreted by the authors as due to photoelectric absorption by
neutral iron close to the source. The lack of any iron L-shell
absorption in the object, which has a strong soft X-ray excess,
implies that the absorber only partially cover the source. This
requires an unusual geometry for the source and absorber since the
source is rapidly variable. However, a different interpretation of the
spectral shape is also possible. As pointed out by Boller et al
(2002), the blue horn of a strong and relativistically blurred iron
K$_\alpha$ line, combined with the edge of the associated
reflection component, can reproduce the 7~keV
drop. A second observation (hereafter AO2) of 1H\,0707-495, reported
by Gallo et al (2004), shows a similar drop but at a higher energy of
about 7.5~keV in the source rest frame.  The simplest explanation put
forward is that the absorber is outflowing at $0.05c$, which implies
either a prodigious mass and energy loss rate in the outflow or a very
clumpy, local, absorber.

The alternative model in which the X-ray spectrum is reflection
dominated was discussed in more detail by Fabian et al (2002b) for the
GT observation. A corrugated disc was invoked to account for the X-ray
reflection dominating the observed emission. A reflection spectrum
here means the back-scattered, fluorescent, recombination and other
emission produced when an X-ray continuum irradiates a slab of matter
(see e.g. Ross \& Fabian 1993; Zycki et al 1994; Nayakshin, Kazanas \&
Kallman 2000; Ross, Fabian \& Ballantyne 2002; Rozanska et al
2002). The continuum source was inferred to be mostly hidden from
direct view in a corrugation, but its reflection observable from the
walls. If the reflecting material is the inner parts of an accretion
disc then the reflection spectrum is blurred by Doppler and
gravitational redshift effects. The sharp 7~keV drop in 1H\,0707-495
is then the blue edge of the relativistically-blurred iron emission
line. The shift in energy of the edge during the second XMM-Newton
observation can then be due to a change in ionisation or geometry of
the irradiated material.

Relativistically-blurred, reflection-dominated spectra have been seen
e.g. in \mcg. There the spectral variability in normal flux states can be
attributed to a two-component model consisting of a power-law
continuum, with a fixed spectral index but changing flux, and an
almost constant reflection component. The model requires that the
continuum source is anisotropic, as viewed by us. This can be
explained by the strong gravitational light bending (see e.g.
Martocchia \& Matt 1996) expected close to the black hole (spectral
fits to the broad iron line in \mcg\ imply that the inner radius of
the accretion disc where much of the reflection originates is about 2
gravitational radii or less, Wilms et al 2001; Fabian et al 2002a).

The variability of the two components in \mcg\ can be explained by a
recently proposed light bending model (Miniutti et al 2003; Miniutti
\& Fabian 2004).  Computations of the change in the observed power-law
and reflection components as the continuum source moves in height
above the black hole are consistent with the otherwise puzzling
behaviour of the broad iron line in this object. Gravitational light
bending naturally produces reflection--dominated and relativistically
blurred spectra during low flux states and thus provides a possible
alternative scenario for the GT XMM-Newton observation of \h. A major
prediction of the model is that reflection is correlated with the
power law continuum only at very low flux levels, while at higher
fluxes it is almost constant despite large continuum variability. The
light bending model also explains the behaviour of the reflection and
iron line in the Galactic Black Hole Candidate XTE\,J1650-500
(Miniutti, Fabian \& Miller 2004; Rossi et al. 2004).

Here, we apply the reflection model to the AO2 observation of \h\ to
investigate whether reflection-dominated emission can account for the
whole XMM-Newton spectrum and its variations. Our purpose here is not
to carry out a detailed study but to check whether such a model gives
a reasonable fit to the data for plausible parameters.

\section{An emission line in the spectrum of \h?}

\begin{figure}
\includegraphics[width=0.4\textwidth,angle=-90]{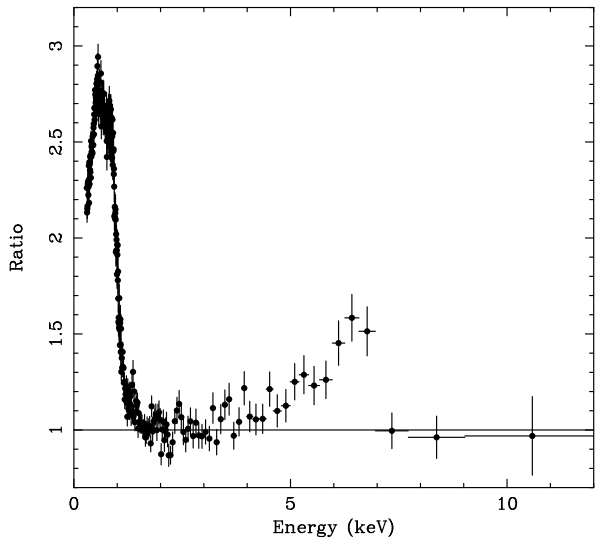}
\caption{Ratio of the XMM-Newton pn spectrum of \h\ to the power-law
which fits best between 1.5--2~keV and 7.5--12~keV. A broad iron line
and a steep soft excess are evident. Data have been rebinned for
visual clarity.
} 
\end{figure}

We first look at the AO2 observation of \h\ from the point of view of
an iron emission line. The XMM-Newton spectrum used here is the EPIC
pn spectrum of Gallo et al (2004), where details of the data reduction
can be found. A power-law was fit to the data in the bands 1.5--2~keV
and 7.5--12~keV. The ratio of the whole 0.3--12~keV dataset to that
power-law is shown in Fig.~1. A large skewed emission feature, similar
to that found in \mcg\ (Tanaka et al 1995; Wilms et al 2001; Fabian et
al 2002a) is evident between about 4 and 7~keV, and a steep soft excess
below 1~keV.

The data between 1.5 and 12~keV were then modelled with a power-law and
a relativistic emission line (Laor 1991). A good fit is obtained (with
$\chi^2 = 365$ for 334 dof, see Fig.~2) for $\Gamma = 2.8$. 
Notice that no extreme
inclination is required to reproduce the spectral drop ($i \simeq
40-50^\circ$). The strength of the line (equivalent width of 1.8~keV) and
its rest frame energy (6.7~keV) imply that we are dealing with an
ionised reflector. The reflection continuum is then
complex and needs to be modelled properly. 
\begin{figure}
\includegraphics[width=0.28\textwidth,angle=-90]{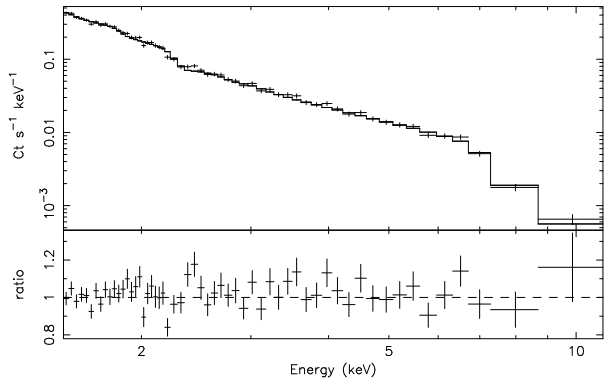}
\caption{The best-fitting power-law plus Laor (1991) line model is shown
in the 2--12~keV band plotted on the spectrum of \h. Data have been
rebinned for visual clarity.
} 
\end{figure}

\begin{figure}
\includegraphics[width=0.28\textwidth,angle=-90]{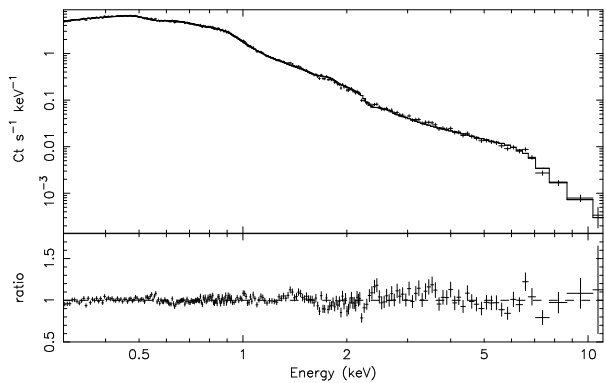}
\caption{The best-fitting (blurred) ionised reflection plus power-law model
plotted on the spectrum of \h. Simple photoelectric absorption has
also been applied. The model fits the data reasonably well over the
entire observed energy band. Data have been rebinned for visual clarity.
} 
\end{figure}

\section{Reflection dominated spectra}

We have therefore fitted Ross \& Fabian (1993) ionised reflection
models (from a set computed by D.R. Ballantyne) to the whole
0.3--12~keV observable band. The line strength and past work on \h\ 
(Fabian et al 2002b) suggest using super--solar abundance models.
Reasonable agreement is obtained (Fig.~3) with a model which is
relativistically blurred, has 3 times solar abundance, and photon
index $\Gamma=3$ (which is the maximum in the available grid) giving
$\chi^2 /dof=760/575$.  The ionisation parameter of the reflection
model is $\log\xi=2.8$.  This is not as good a fit as Gallo et al
(2004) obtained with a partial-covering model ($\chi^2 /dof
=655/562$), but that had 5 separate components (apart from cold
absorption), many of which operate on separate parts of the spectrum,
whereas the reflection model has only two components (power-law and
reflection), both of which are broadband.  In Fig.~4, we show the
(blurred) reflection plus power law model used to describe the data.
No other components were added to obtain the result shown in Fig.~3.

Much of the difference ($\Delta \chi^2=105$) in the quality of the
fits between this model and the partial covering one is due to the
soft X-ray band. We have a simple photoelectric absorption component
({\tt{WABS}}) which requires a larger column density ($1.4\times
10^{21}$~atoms~cm$^{-2}$) than the Galactic value ($5.8\times
10^{20}$~atoms~cm$^{-2}$, see e.g. Gallo et al 2004). If the excess
absorption is instead allowed to be mild edges such as in a warm
absorber, then most of the $\chi^2$ difference is eliminated ($\chi^2
/dof =685/572$). The same good quality can be obtained by considering
instead a multiple reflection model such as that presented by Fabian
et al. (2002b) for the GT observation. It is not our purpose here to
discuss in detail the possible parameterisations of the spectrum. We
merely stress here that a reasonable fit can be obtained even with the
simplest possible reflection model and that the reduced $\chi^2$
values in the 2--10~keV band for the partial-covering and reflection
models are almost identical.
\begin{figure}
\includegraphics[width=0.43\textwidth,angle=-90]{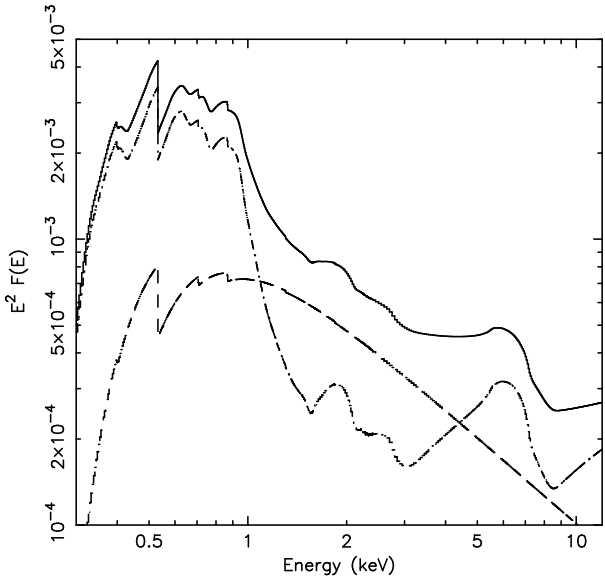}
\caption{The (blurred) ionised reflection plus power law model used in 
 the fit shown in Fig.~3.  }
\end{figure}

The intrinsic unblurred reflection spectrum is shown in Fig.~5. A
large iron-K emission line and edge dominate the spectrum around 7~keV
and a steep soft excess occurs below 1 keV (see also Fig.~4 which
includes blurring and photoelectric absorption). The bump in the
blurred spectrum just below 1~keV, modelled as a large Gaussian
emission line by Gallo et al (2004), is due to a complex of iron-L
lines. The level of the reflection component relative to the power law
is about 9 times that expected for irradiation of a flat disc. This
implies that much more primary radiation is illuminating the disc than
reaching the observer at infinity.

The
relativistic blurring is accomplished using a Kerr kernel (Laor 1991)
with a radial emissivity index of $q\simeq 5.1$ (where the emissivity is has
the radial dependence $r^{-q}$) from $r_{\rm{in}} \simeq 2.25$ to
$r_{\rm{out}} = 100$ gravitational radii, and disc inclination of
about $50^\circ$. The disc inclination is different from that obtained
in the previously published analysis of the GT observation ($i <
30^\circ$, see Fabian et al 2002b). We shall return on this issue later
in Section 4.
\begin{figure}
\includegraphics[width=0.4\textwidth,angle=-90]{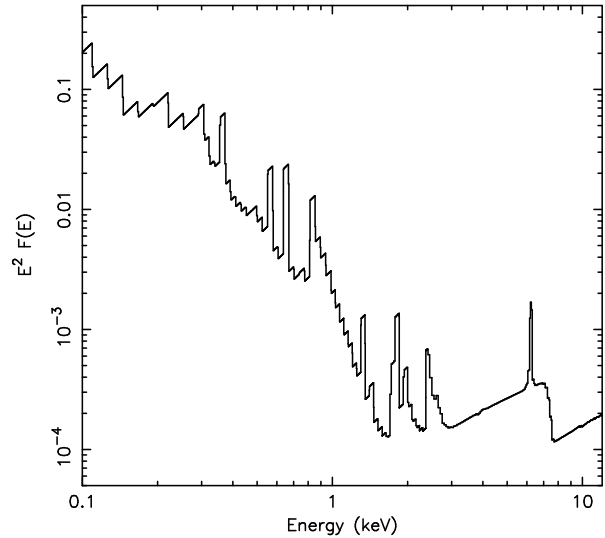}
\caption{The unblurred best-fitting reflection model. 
The abundances are 3 times Solar and
  $\xi=10^{2.8}$. The irradiated gas is assumed to be of constant
  density with no intrinsic emission.}
\end{figure}

We have investigated whether models with higher abundances improve on
the fit of the second (AO2) observation. At 7 times solar abundances the
drop is sharper (see Fig.~6; the spectrum was fitted above 2~keV and is
shown above 5~keV), but overproduces emission below about 0.7~keV.
This is probably because the oxygen abundance is now too high. The
data appear to require a high iron abundance but smaller oxygen
abundance. A more extensive grid of variable abundance reflection
models is required to pursue this issue further (and would likely reduce
the low energy residuals in our present analysis). A range of ionisation
parameters may also be needed to match the range of disc radii, so
leading to a highly multi-parameter solution which is beyond the scope
of this paper. 
\begin{figure}
\includegraphics[width=0.4\textwidth,angle=-90]{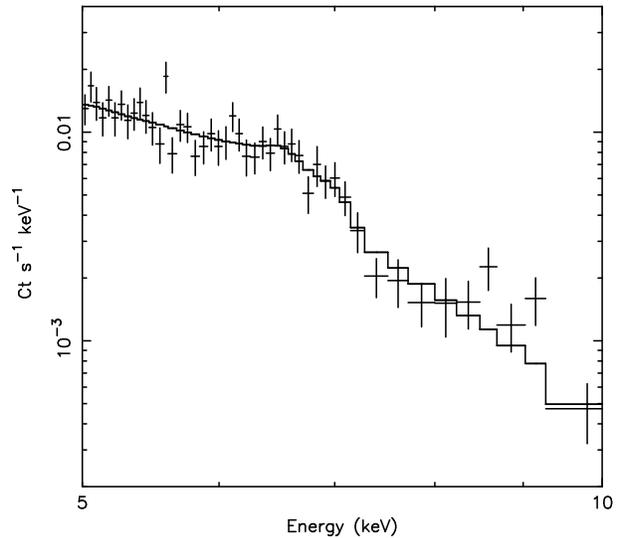}
\caption{The higher energy spectrum fitted with a reflection model which has
abundances at 7 times solar. The large drop in the model appears to fit the 
data above 7~keV better. }
\end{figure}

\section{Comparison between the GT and AO2 observations}

As already mentioned, \h\ has been observed twice by XMM-Newton.
During the first GT observation the count rate was significantly
lower  than during the AO2 observation (see Fig.~5
in Gallo et al 2004). In both observation a spectral flattening above
$\sim$~3~keV is evident and according to our interpretation it is related
to the presence of a reflection component including a broad iron
emission line. The most remarkable difference between the
two observations is that the drop energy, when modelled by a simple
edge, shifted from $E \simeq 7$~keV in the GT observation to $E \simeq
7.5$~keV in the AO2 observation. Moreover, the edge depth is
considerably smaller during the higher flux AO2 observation (about a
factor 2), while the width is narrow at both epochs.

The reflection--dominated model for the GT observation was presented
by Fabian et al. (2002b). The broadband spectrum was described by a
multiple reflector model (all components being blurred), and no power
law continuum was required, producing a spectrum completely dominated
by reflection. The presence of different reflection components with
different ionisation parameters most
likely reproduces the radial distribution of ionisation 
on the disc, and does not allow to easily derive the average ionisation
state; however, a value around $\log\xi = 2$ seems to be a reasonable
estimate for the GT observation, while we measure $\log\xi = 2.8$ in
the AO2 data. 
\begin{figure}
\includegraphics[width=0.28\textwidth,angle=-90]{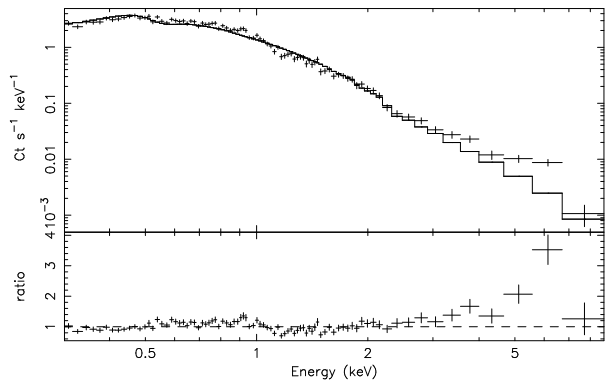}
\caption{ The difference spectrum, made by subtracting the high and
low spectra,  with the best fitting power-law plus {\tt{WABS}} model. }
\end{figure}

The best--fit model for the GT observation was obtained by fixing the
emissivity index of the relativistic blurring model to $q=3$ and the
disc inclination was derived to be about $20^\circ$. The inclination
we are measuring here is around $50^\circ$ and seems to contradict the
GT analysis. However, we re--fitted the best--fit model of Fabian et
al. (2002b) for the GT observation letting the emissivity index free
to vary during the fit.  We find that a common solution does exists
for the two observations with inclination $\sim 50^\circ$. The new fit
for the GT observation is of the same statistical quality of that
presented in Fabian et al.  (2002b) with $\chi^2 / dof = 288/303$. The
main difference between the two observation is that the emissivity
index is much steeper during the GT ($q \simeq 7$) than the AO2
observation ($q \simeq 5$). A flatter emissivity profile reduces the
effect of gravitational redshift on the sharp reflection features
(emission line and edge) so that the edge is observed at slightly
higher energy during the AO2. It is therefore a combination of higher
ionisation parameter and flatter emissivity profiles that is
responsible for the edge energy shift from the GT to the AO2
observations. As mentioned, the other major difference between the two
observations is that the spectrum is completely dominated by
reflection during the GT observation, while a power law continuum
becomes visible during the AO2 (compare Fig.~4 of the present paper
with Fig.~5 in Fabian et al 2002b) during the AO2.

\subsection{Source variability}

\h\ varied dramatically during the second XMM-Newton observation.
Spectra were obtained when the source exceeded 4.5 cps and was less
than 3.5cps (see Fig.~1 in Gallo et al 2004 for the broadband light
curve). The difference between them is shown in Fig.~7 when fitted
with a power law and cold absorption. It is close to a power-law of
index $3.55$, but shows a bump at iron-K energies. The source is
therefore not identical in behaviour to \mcg\ where a difference
spectrum mostly shows the power-law component. In \h\ the reflection
component must vary too. In particular, since a bump is seen in the
4--7~keV band, the broad iron line must be stronger at high than low
flux. We have fitted the high and low-flux spectra separately allowing
only the normalisations of the power law and reflection components to
vary.  Good fits with a reduced $\chi^2\simeq 1$ were found for both.
The reflection normalisation changes by a factor of 1.7 while the
power-law one by a factor of 2.8 (a factor $\approx$~1.6 between the
two). The variability is therefore dominated by changes in the flux of
the power-law continuum, but the reflection component varies as well.

As shown by Gallo et al (2004), the fractional variability amplitude
(F$_{\rm{var}}$) in the AO2 observation is larger (again, by about a factor
$\approx$~1.6) in the intermediate than in the soft
(~0.3--1~keV~) and hard (~5--12~keV~) band. Moreover, F$_{\rm{var}}$
is roughly the same in the soft and hard band, where according to our
spectral analysis (see Fig.~4) ionised reflection dominates.
Therefore, in the reflection--dominated model we present here, the
F$_{\rm{var}}$ behaviour as a function of energy is easily understood.
The 1--5~keV band is dominated by the power law continuum which
contributes the most to the source the variability. The soft and hard
bands do exhibit the same F$_{\rm{var}}$ because they are both
dominated by ionised reflection from the accretion disc, which is less
variable than the power law continuum. During the first GT XMM-Newton
observation of \h, F$_{\rm{var}}$ was found to be constant (see Fig.~4
of Boller et al 2002), as opposed to the AO2 results in Gallo et al
(2004). As mentioned, spectral analysis with the reflection--dominated
model strongly suggests a completely reflection--dominated spectrum
during the GT observation. Notice that even if not formally required,
spectral analysis cannot exclude the presence of a faint power law
component in the GT spectrum. The constant F$_{\rm{var}}$ is then the
result of a dominant broadband spectral component in the overall
0.3-12~keV spectrum, i.e.  ionised reflection from the disc, or
indicates that if a power law is present, it varies with the same
amplitude as the reflection component.
\begin{figure}
\includegraphics[width=0.33\textwidth,angle=-90]{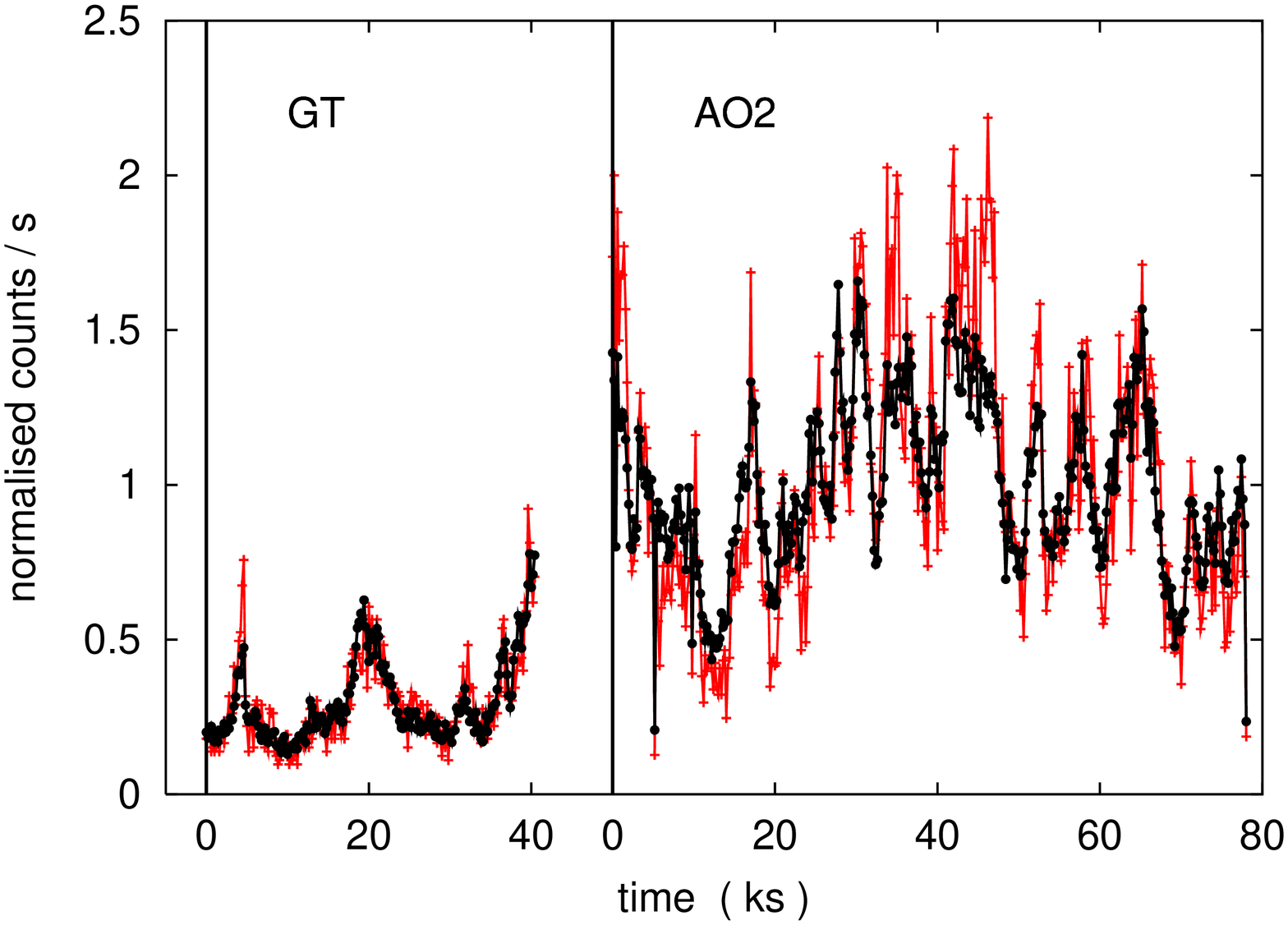}
      \vspace{0.2cm}
\includegraphics[width=0.33\textwidth,angle=-90]{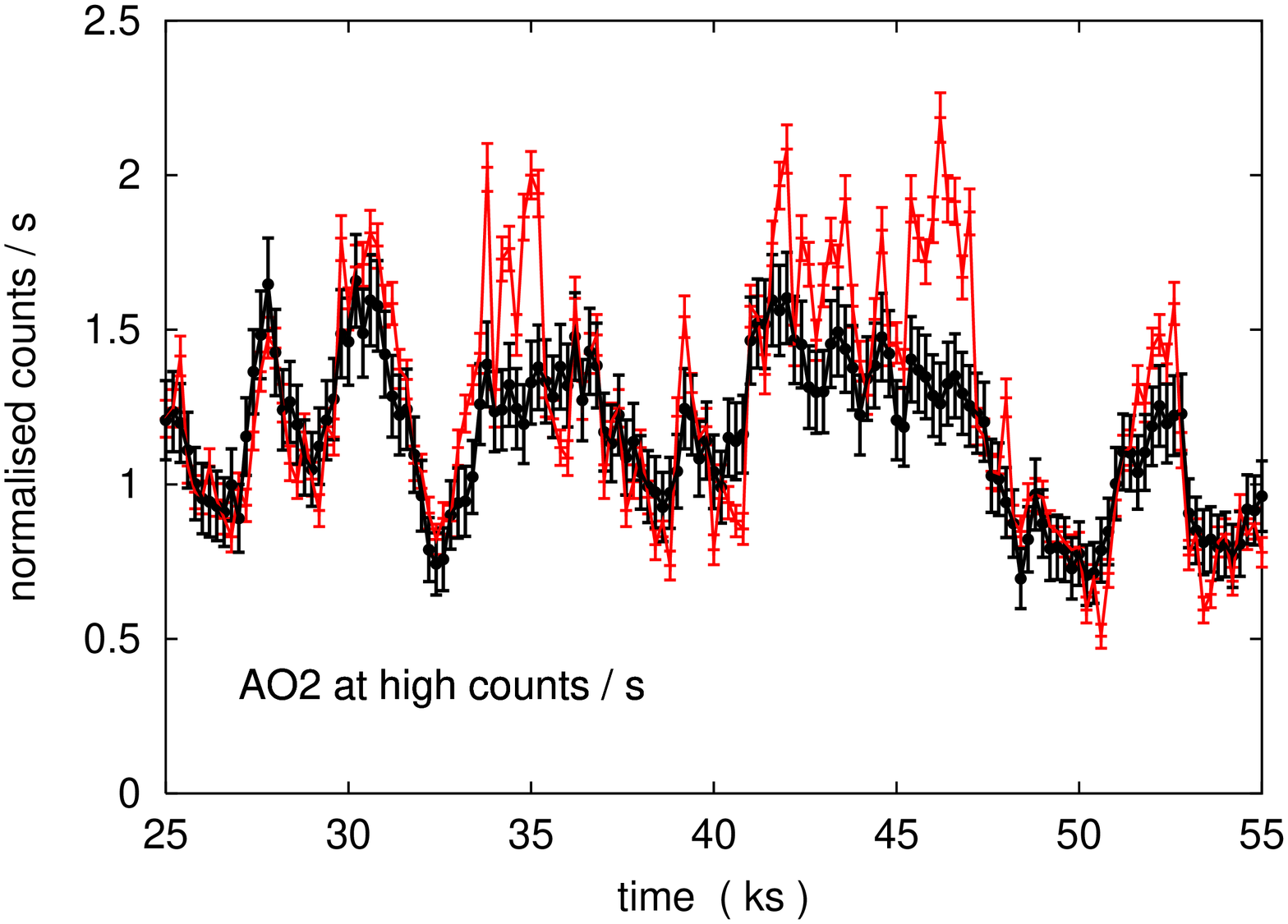}
\caption{ {\emph{Top:}} normalised light curves 
  in the 0.3-1~keV (black solid circles) and 1--5~keV (red open
  triangles) bands for the GT and AO2 observations (200-seconds
  binning). {\emph{Bottom:} a zoom in the high count rate region for
    the AO2 observation. Notice that the 0.3--1~keV light curve tracks
    very well the 1--5~keV but appears to saturate as the count rate
    increases. In this case, error bars are also shown.}}
\end{figure}

\subsection{Energy--selected light curves}

If the reflection-dominated spectrum is produced by strong light
bending effects, we expect the reflection--dominated component (RDC)
to respond to the power law component (PLC) variability only at low
flux levels and to vary much less than the power law at higher
fluxes (see e.g.  Miniutti \& Fabian 2004). This predicted behaviour
should be visible in energy-selected light curves if, as the previous
discussion indicates, reflection dominates below $\sim 1$~keV and
above $\sim 5$~keV and the power law continuum provides the largest
contribution in the intervening band (see Fig.~4).  Guided by our
spectral analysis the 0.3--1~keV and 1--5~keV bands were selected as
representative of the two main spectral components of \h. 

In the top panel of Fig.~8 we show the $200$-s binned light curve in
the $0.3-1$~keV (black) and $1-5$~keV (red) bands for the GT and AO2
observations.  The light curves were normalised to their mean count
rate and the GT light curves were then rescaled by the ratio of
$0.3-5$~keV count rates in the two observations to account for the
lower count rate in GT observation (a factor $\sim 3.4$).  It is clear
from the figure that there is a good correlation between the two
bands in both observations. However one major difference is seen at
the highest count rates during AO2: the 0.3--1~keV light curve does
not reach the high count rate level of the 1-5~keV. The bottom panel
of Fig.~8 shows a close-up of the AO2 light curves in the highly
variable period between $25-55$~ks.  Clearly the two light curves are
more divergent at higher fluxes. However, this can be explained as due
to the `rms-flux' correlation known to be present in Seyfert galaxy
X-ray variations (Uttley \& M$^{\rm c}$Hardy 2001; Vaughan, Fabian \&
Nandra 2003).  If the two light curves follow this relation then
differences in Fig.~8 can be explained largely by a more variable
    $1-5$~keV than $0.3-1$~keV light curve. The more variable
($1-5$~keV) light curve is thereby more `non-linear' and the two will
naturally diverge at higher fluxes (see Uttley, M$^{\rm c}$Hardy \&
Vaughan 2004, in prep).

As a further investigation of the nature of the spectral variability
in \h, flux-flux plots were examined (see Taylor Uttley \& M$^{\rm
c}$Hardy 2003; Vaughan \& Fabian 2004). Light curves were extracted in
six bands and binned to $500$-s resolution (ensuring a minimum of
$>20$ ct/bin). The flux in each band was plotted against the flux in
the $1-2$~keV band. (This band was chosen as the comparison band as it
is dominated by the variable power law and, as confirmed by the
subsequent analysis, contains the smallest fraction of constant
component of any of the six energy bands.) The data were then binned as
a function of $1-2$~keV flux (such that there are $>10$ fluxes per
bin). Fig.~9 (top panel) shows one of the resulting flux-flux plots.
Following the above discussion, examining the correlation between
variations in different bands should reveal the RDC versus PLC
relationship.
\begin{figure}
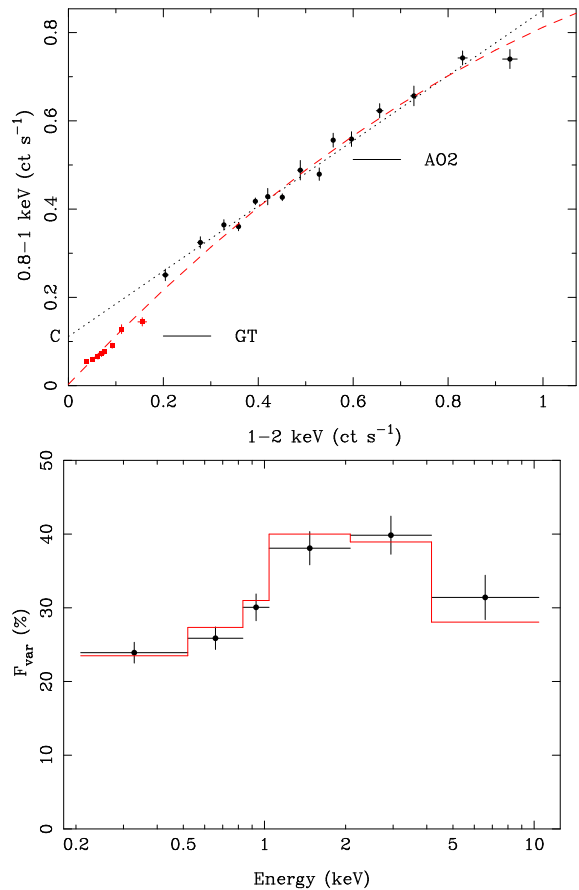

\includegraphics[width=0.33\textwidth,angle=-90]{fig10.ps} 
      \vspace{0.2cm}
\includegraphics[width=0.33\textwidth,angle=-90]{fig11.ps}
\caption{ {\emph{Top:}} 
Flux-flux plot for a soft band ($0.8-1.0$~keV) light curve
against the $1-2$~keV light curve. Clearly during the (higher
flux) AO2 observation the relation between the fluxes is 
quite linear (dotted curve), with a constant offset (marked `C'). 
Including the (lower flux) GT data shows a significant deviation
for the linear relation at low fluxes, approximated better
by a quadratic model (dashed curve).
{\emph{Bottom:}} 
Rms spectrum of \h\ calculated using $500$-s binned light curves. 
As reported by Gallo et al. (2004) the variability amplitude is
energy-dependent and peaks in the $1-4$~keV range. The histogram
shows a model rms spectrum produced assuming the spectrum comprises
only a variable power law and a constant component as defined
by the offsets of the flux-flux relations (the `C' shown above).
}
\end{figure}

Considering the AO2 data alone, a linear model gives a reasonable
match to the data (although not acceptable in a formal sense, with
$\chi^2 = 33.08 / 13 ~ dof$). The value of the constant (marked `C' in
the figure) gives an indication of the contribution to the spectrum of
any truly constant emission, which must be stronger in the $0.8-1$~keV
band than the $1-2$~keV band.  The approximate linearity implies that
the variable component of the spectrum has a flux-independent spectral
shape. Following the analysis of Taylor et al. (2003) and Vaughan \&
Fabian (2004) the value of the constant C was recorded as a function
of energy band. This gives an estimate of the spectrum of the constant
component that can be used to model the energy-dependent variability
amplitude as described in section 5.3 of Vaughan \& Fabian (2004).
Assuming only two spectral components, one variable (with constant
spectral shape) and one constant in flux (as derived using the above
method) a model for the rms spectrum was produced. Fig.~9 (bottom
panel) shows the comparison between the model and the measured rms
spectrum. The agreement is rather good, indicating that such a simple
description of the amplitude-dependent variability is a reasonable
first-approximation.  The spectrum of the constant component is in a
sense the `opposite' of the rms spectrum, with strong emission below
$1$~keV and above $5$~keV, which suppresses the variations.
Qualitatively, the constant component resembles the ionised reflection
spectrum (Fig.~4). However, the RDC cannot be completely constant. This is
because the RDC is a factor $\sim 10$ brighter than the PLC below $1$
keV, and if this were completely constant it would suppress the
variability in this band to a much greater degree than actually
observed. Therefore, as also indicated by spectral analysis (see
Section 4.1), the RDC must vary too but with smaller amplitude than the
PLC.

The situation is undoubtedly more complex than that described above.
Firstly, for the rms-flux relation to account for the
divergence in the light curves at high fluxes there must be
differences in the amplitude of the intrinsically variable
spectrum (i.e. after accounting for any constant)  as a
function of energy. Secondly, considering the GT
observation, which was obtained during a
period of much lower flux, reveals that the approximate linearity of
the flux-flux relation cannot extend to very low fluxes (Fig.~9).

This can be explained (qualitatively at least) in terms of the light
bending model as follows. At intermediate fluxes (e.g. the AO2
observation) the RDC is largely unresponsive to changes in the
observed PLC flux (regime II of Miniutti \& Fabian 2004) whereas at
low fluxes the two become correlated (regime I).  Furthermore,
the prediction of this model is that the RDC and PLC become tightly
coupled at low fluxes, which would result in a much flatter rms
spectrum, exactly as observed during the GT observation (Fig.~4 of
Boller et al. 2002).  For \h, the flux-flux relations including the
AO2 and GT data are better fitted with quadratic model to account for
the low-flux curvature. Notice that a
similar quadratic relationship is found for the (broad) iron line flux
versus the continuum flux in the Galactic black hole candidate
XTE~J1650-500 (Rossi et al 2004; see also Miniutti, Fabian \& Miller
2004). Indeed, the light bending model predicts that plotting the disc
RDC (including the broad iron line) against the primary PLC results in
a relationship which is much better fitted by a quadratic than a
linear curve (see Fig.~2 in Miniutti \& Fabian 2004).
   
Both the GT and AO2 observations of \h\ match well the predictions of
the light bending model. In this model, low flux states (GT
observation) are associated with strong light bending which naturally
produces reflection--dominated spectra and very steep emissivity
profiles on the disc. During higher flux states (AO2 observation)
light bending effects are slightly reduced, the emissivity is
(relatively) flatter, and the power law continuum becomes detectable.

If our interpretation is correct, we expect that observations in
higher flux states than AO2 will result in a larger contribution of
the power law continuum above about $1$~keV leading to a further
reduction of the depth of the spectral drop at $7-8$~keV (consider
Fig.~4 and shift the PLC at a higher level keeping the RDC almost
constant). During such a `high flux state' the variability should be
dominated by the PLC producing much larger variability in hard than
soft bands. \h\ was in fact observed by {\it{Chandra}} a few months
later than the GT {\it{XMM--Newton}} observation, and the source flux
was found to be $\sim 10$ times larger than during the GT (and
therefore larger than during AO2 as well), as reported by Leighly et
al. (2002). Leighly et al. pointed out the lack of any spectacular
spectral drop around $7-8$~keV in the {\it{Chandra}} high flux
observation in contrast with both {\it{XMM--Newton}} spectra.
Moreover, the authors found that the hard X--rays above about 1~keV
are much more variable than the soft ones. Both observational results
agree very well with the predictions of the light bending model.

\section{Discussion}

Remarkably, a relativistically blurred reflection-dominated model
gives a reasonable fit to the new XMM-Newton spectrum of \h\ over the
entire observed energy band.  The parameters are similar to those of
\mcg\ and NGC4051 in their lowest state, but with a steeper photon
index.

The result on \h\ is consistent with fits to the earlier GT data
(Boller et al 2002; Fabian et al 2002b), except that the photon index
of the illuminating power-law was then slightly flatter at
$\Gamma\sim2.5$, rather than $\Gamma\sim 3$ found here. The edge energy
shifted from 7~keV to 7.5~keV in two years. The shift 
is not principally due to a change in ionisation
state. Although the best fits shows an increase in $\log \xi$ from 2
to 2.8, the energy of the iron line remains consistent with the
helium-like emission line. We have refitted the first dataset (GT
observation) searching for solutions with ionised reflection from the
accretion disc with same inner/outer disc radius and inclination as
observed in this second observation. Common solutions do exist and
require that the emissivity index is larger in the GT than in the AO2
observation.  This has the effect of emphasising the inner radii of
the disc and so increasing the effect of gravitational redshift,
explaining why the sharp drop occurs at a lower energy in that
dataset. 

A power law component (not visible during the GT observation)
dominates the 1--5~keV band in the (higher flux) AO2 observation. The
spectral model also accounts for the lack of any evident narrow iron
K$_\alpha$ emission line (the line being relativistically blurred).
and requires no absorption feature in the 1-1.5~keV band.  Several
NLS1 objects with the strongest soft excess exhibit an apparent
absorption feature near 1~keV (e.g. Leighly 1997). The model we
propose here for \h\ does require no absorption feature in the
1-1.5~keV band, but has instead line emission just below 1~keV. It is
therefore possible that the spectra of other NLS1s around 1~keV can be
explained by ionised reflection as well.

Strong gravitational light bending can explain the behaviour of \h\ 
and possibly other NLS1, as discussed by Miniutti \& Fabian (2004).
This seems to be the simplest explanation of the change in the
spectrum. The model predicts that 
the lowest flux state occurs when the continuum
source is closest to the centre of the disc, the power law
contribution is depressed and reflection--dominated spectra which are
relativistically blurred with steep emissivity profiles are produced.
Multiple reflection associated with returning radiation or modelling a
radial profile of the ionisation parameter may provide the most
interesting spectral parametrisation (Ross, Fabian \& Ballantyne
2002). The observed rapid variability can be associated with changes in
height of the continuum source above the black hole. This produces a
correlation between the RDC from the disc and the PLC at low flux (GT
observation), while at higher flux (AO2 observation), the model
predicts a variable PLC and an almost constant RDC. 

The fractional
variability amplitude behaviour in both the first and the second
XMM-Newton observations and the analysis of energy selected light
curves supports this interpretation. According to our spectral
parametrisation, the soft band is dominated by ionised disc reflection
and the hard one by the power law continuum. The flux--flux plots
analysis demonstrates that during the AO2 observation the spectral
variability can be understood in terms of a variable PLC, and a more
constant RDC, in excellent agreement with the light bending model. The
observed rms spectrum is very well reproduced by this simple model and
the derived spectrum of the constant component closely resembles that
of ionised disc reflection with strong emission below $1$~keV and
above $5$~keV.

As pointed out by Miniutti \& Fabian (2004) the inner disc emission is
preferentially beamed along the equatorial plane thus illuminating the
outer regions of the disc. This low velocity, high density material
would then produce intermediate and low ionisation emission lines
which are observed as narrow, as opposed to the blueshifted, broad
high ionisation lines likely produced in a lower density and high
velocity wind (Leighly \& Moore 2004, Leighly 2004a). The high
ionisation lines then appear predominantly blueshifted since the disc
itself prevents us from seeing the outflow on the other side. Since
the disc emission (here ionised reflection) has a very steep spectral
shape and peaks below $1$~keV, only a minor fraction is available for
iron fluorescence above 7~keV. The associated narrow iron line is
therefore weak and, if present, would be difficult to detect above the
level of the much stronger inner disc emission dominated by the broad
line in the 5--7.5~keV band.

The present work suggests that a much deeper study of both datasets
with expanded grids of reflection models will be fruitful. The results
so far indicate that the emission from \h, and likely other NLS1,
originates from within a few (10 or so) gravitational radii from
the black hole. NLS1 may be among the best sources with which to probe
the extreme gravity at a few gravitational radii around the black
hole.

\section*{Acknowledgements}
Based on observations obtained with {\it{XMM--Newton}}, an ESA science
mission with instruments and contributions directly funded by ESA
Member States and the USA (NASA). ACF thanks the Royal Society for
support. GM thanks the PPARC for support.

\label{lastpage}
\end{document}